\documentstyle[aps,eqsecnum,preprint,psfig]{revtex}
\newcommand{\Eref}[1]{Eq.~(\ref{#1})}                                        %
\newcommand{\Sref}[1]{Sec.~\ref{#1}}                                         %
\newcommand{\Fref}[1]{Fig.~\ref{#1}}                                         %
\newcommand{\Rref}[1]{Ref.~\cite{#1}}                                        %
\newcommand{\etal}{{\it et al.\/}}                                           %
\newcommand{\e}{{\rm e}}                                                     %
\renewcommand{\i}{{\rm i}}                                                   %
\newcommand{\ba}{\begin{array}}                                              %
\newcommand{\ea}{\end{array}}                                                %
\newcommand{\eq}[2]{\begin{equation}#2\label{#1}\end{equation}}              %
\newcommand{\eqn}[1]{\begin{eqnarray}#1\end{eqnarray}}                       %
\newcommand{\nn}{\nonumber}                                                  %
\newcommand{\suma}{\sum\limits}                                              %
\newcommand{\eps}{\varepsilon}
\newcommand{\ec}{\varepsilon_{\rm c}}
\newcommand{\eb}{\varepsilon_{\rm b}}
\newcommand{\ew}{\varepsilon_{\rm w}}
\newcommand{\NCCO}{{\rm Nd}_{2-x}{\rm Ce}_x {\rm CuO}_{4-\delta}}
\tighten
\begin{document}
\draft
\preprint{Bulgarian Journal of Physics {\bf 24}, Nos 3/4 (1997) 114-125}
\title{
Superconducting gap anisotropy within the framework of a simple exchange model
for layered cuprates.\\ The theory of HTSC}
\author{T. M. Mishonov
\thanks{Fax: (+359 2) 96 252 76;
Tel.: 6 256 652;
E-mail: mishon@phys.uni-sofia.bg},
A. A. Donkov, R. K. Koleva\\
{\it Department of Theoretical Physics, Faculty of Physics, University of
Sofia\\ 5 J. Bourchier Blvd., 1164 Sofia, Bulgaria}\\
E. S. Penev\\
{\it Institute of Catalysis, Bulgarian Academy of Sciences\\
Acad. G. Bonchev Str, bl. 11, 1113 Sofia, Bulgaria}
}
\date{Received 20 October 1996}
\maketitle
\begin{abstract}
The oxygen O$2p_{\sigma}$ and copper Cu$4s$ and Cu$3d_{x^2-y^2}$ orbitals
are involved in a simple LCAO model for determination of the conduction
band and the oxygen-oxygen hopping is considered as a small parameter with
respect to the transition amplitude between nearest neighbours. The
traditional Cooper pairing is obtained by taking into account the
double-electron exchange between the nearest neighbours within the
two-dimensional CuO$_2$ plane. The equation for the superconducting gap is
obtained as a result of the standard BCS treatment. It is shown that
the order parameter could have either $s$-type or
$d$-type symmetry depending on
the ratio between the transition amplitudes. This model allows
understanding the experiments reporting a $\pi$-shift of the Josephson phase
indicative for a $d$-type gap symmetry as well as the observed $s$-type
in the case of strongly irradiated samples.
\end{abstract}


\section{Introduction}                                                     %
\label{sec:intro}                                                          %

The discovery of high-$T_c$ superconductivity~\cite{Bednorz} has brought
significant interest into this field and triggered many intense investigations
during the last decade. In the centre of them is the question about the
determination of the basic parameter characterizing this phenomena
-- the superconducting order parameter. Recent experiments on
angle-resolved photoemission spectroscopy (ARPES) gave fast increase of the
quantitative results for the Fermi surface and for the type of the
angular dependence of the order parameter as well. As a result of these
studies, for YBa$_2$Cu$_3$O$_7$ and Bi$_2$Ca$_2$SrCu$_2$O$_8$ symmetry of the
type $\cos p_x-\cos p_y$ is often assumed. This assumption has been
independently confirmed by the experiments on Josephson
junctions for YBa$_2$Cu$_3$O$_7$~\cite{Mathai}. On the other hand, deviation
from the simple $d$-case was observed in the experiments with strongly
irradiated samples~\cite{Radtke}.  As suggested by Abrikosov~\cite{Abrikosov}
and Pokrovsky and Pokrovsky~\cite{Pokrovsky} that could be realized by
the reduction of the $d$-channel and domination of the $s$-part of the
electron-electron interaction. The $\NCCO$, for example,
has similar symmetry  of the superconducting gap.

The purpose of this paper is to derive analytical expression for the
interaction involved in the standard equation for the superconducting gap
by successive BCS treatment of the double-electron exchange between the nearest
neighbours (NN) and the next nearest neighbours (NNN) in the CuO$_2$ plane.
The matrix elements of the interaction, being a sum of $s$- and $d$-symmetry
terms, and the limit cases leading to simple $s$- or $d$-type gap are
discussed as well.

The eigenfunctions and eigenvalues for the conduction band found by
solving of Schr\"odinger equation are used to obtain the momentum
representation of the {\it superexchange interaction}. The successive BCS
scheme applied to the latter leads to equation for the BCS gap. The
interpretation of the exact result in the limit cases of strong hole and
electron doping is discussed in \Sref{sec:Discussion}. In conclusion the
fitting of our results to the recent experimental data is considered.

\section{Model}                                                           %
\label{sec:Model}                                                         %

Following the ideas of quantum chemistry~\cite{Harrison} we shall use a
tight-binding (TB) method to obtain the electronic band structure of layered
cuprates. To this purpose we consider the atomic orbitals related to the
Cu$4s,$ Cu$3d_{x^2-y^2},$ O$2p_{\sigma}$ states. Denoting with ${\bf R}_x,$
${\bf R}_y,$ ${\bf R}_{\rm Cu}$ the positions of O$_x,$ O$_y,$ and Cu atoms in
the CuO$_2$ plane, with  $a_0$ the in-plane lattice constant and {\bf n} -- the
unit cell index, the wave function within the adopted here liner
combination of atomic orbitals (LCAO) approximation reads as
\eqn{
\psi_{_{\rm LCAO}} ( {\bf r} ) & = & \sum_{{\bf n}, \alpha}
     X_{{\bf n}, \alpha} \psi_{_{{\rm O}2p_x}} ({\bf r}-{\bf
n}a_0-{\bf
     R}_{x}) + Y_{{\bf n}, \alpha} \psi_{_{{\rm O}2p_y}} ({\bf r}-{\bf
     n}a_0-{\bf R}_{y}) \\ \nn
&\ & + S_{{\bf n}, \alpha} \psi_{_{{\rm Cu}4s}} ({\bf r}-{\bf
     n}a_0-{\bf R}_{{\rm Cu}}) + D_{{\bf n}, \alpha} \psi_{_{{\rm
     Cu}3d_{x^2-y^2}}}({\bf r}-{\bf n}a_0-{\bf R}_{{\rm Cu}}),
\label{LCAO}
}
where the coefficients  $X_{{\bf n}, \alpha},$ $Y_{{\bf n}, \alpha},$ $S_{{\bf
n}, \alpha},$ and $D_{{\bf n}, \alpha}$ are the amplitudes for the {\bf n}-th
unit cell.
The building of TB Hamiltonian in the terms of second quantization is
reduced to replacing these amplitudes by creation and annihilation
operators satisfying the anticommutation relations of the type
$\left\{X_{\bf n}, X_{\bf m}^{\dagger} \right\}= \delta_{{\bf n},
{\bf m}},$ $\left\{Y_{\bf n}, S_{\bf m}^{\dagger} \right\}= 0.$
Further introduce the notations $I_s$ for the amplitude of the
transition between the Cu$4s$ and O$2p_{\sigma}$ and $I_d$ -- between
Cu$3d$ and O$2p_{\sigma}$ orbitals. The Hamiltonian giving the band
structure of CuO$_2$ plane, which incorporates the oxygen-oxygen hopping
amplitude $t,$ has the form
\eqn{
   H &=& \sum\limits_{\bf n}\left\{  X_{\bf n}^{\dagger}[
              -t \left(Y_{\bf n}-Y_{x+1,y}-Y_{x,y-1}+Y_{x+1,y-1}
             \right)
              -I_s\left(-S_{\bf n}+S_{x+1,y} \right) \right. \nn\\
  & &\qquad\qquad -I_d\left(-D_{\bf n}+D_{x+1,y} \right)
              + \eps_{2p_x} X_{\bf n} ] \nn\\
  & &\quad\quad + Y_{\bf n}^{\dagger}  [
              -t \left(X_{\bf n}-X_{x-1,y}-X_{x,y+1}+X_{x-1,y+1}
               \right)
              -I_s\left(-S_{\bf n}+S_{x,y+1} \right) \nn \\
  & &\qquad\qquad -I_d\left(D_{\bf n}-D_{x+1,y} \right)
              + \eps_{2p_y} Y_{\bf n}] \label{TB} \\
  & &\quad\quad + S_{\bf n}^{\dagger} [
              -I_s \left(-X_{\bf n} + X_{x-1,y}-Y_{\bf n} + Y_{x,y-1}
              \right)
              + \eps_{4s} S_{\bf n} ] \nn \\
  & &\quad\quad\left. +  D_{\bf n}^{\dagger} [
              -I_d\left(-X_{\bf n} + X_{x-1,y}
              +Y_{\bf n} - Y_{x,y-1}
              \right)
              + \eps_{3d} D_{\bf n}]  \right\}, \nn
}
where the single-site energies of O$2p_x,$ O$2p_y,$ Cu$4s$ and
Cu$3d_{x^2-y^2}$ orbitals are denoted by $\eps_{2p_x},$ $\eps_{2p_y},$
$\eps_{4s}$ and $\varepsilon_{3d}$ respectively; the energy is measured from
the oxygen $2p$ level, i.e. it is assumed $\varepsilon_{2p_x} =
\varepsilon_{2p_y}=\varepsilon_{2p}=0.$ Hence, the energies of the copper
orbitals are $\epsilon_s= \varepsilon_{4s}-\varepsilon_{2p}$ and
$\epsilon_d= \varepsilon_{3d}-\varepsilon_{2p}.$
Now using the Bloch waves
\eq{XYSDn}{\ba{ll}
 X_{{\bf n}, \alpha} = {1\over\sqrt{N}}\suma_{p}\e^{\i{\bf p \cdot n}}
 \left(-\i\e^{\i p_x/2}\right)X_{p, \alpha},      &
 X_{{\bf n}, \alpha}^{\dagger} = {1\over\sqrt{N}}\suma_{p'}\e^{-\i{\bf p'
 \cdot n}}\left(\i\e^{-\i p'_x/2}\right)X_{p', \alpha}^{\dagger},\\  & \\
 Y_{{\bf n}, \alpha} = {1\over\sqrt{N}}\suma_{p}\e^{\i{\bf p \cdot n}}
 \left(-\i\e^{\i p_y/2}\right)Y_{p, \alpha}, &
 Y_{{\bf n}, \alpha}^{\dagger} = {1\over\sqrt{N}}\suma_{p'}\e^{-\i{\bf p'
 \cdot n}}\left(\i\e^{-\i p'_y/2}\right)Y_{p', \alpha}^{\dagger},\\  & \\
 S_{{\bf n}, \alpha} = {1\over\sqrt{N}}\suma_{p}\e^{\i{\bf p \cdot
 n}} S_{p, \alpha}, &
 S_{{\bf n}, \alpha}^{\dagger} = {1\over\sqrt{N}}\suma_{p'}\e^{-\i{\bf p'
 \cdot n}} S_{p', \alpha}^{\dagger},\\                       & \\
 D_{{\bf n}, \alpha} = {1\over\sqrt{N}}\suma_{p}\e^{\i{\bf p \cdot n}}
                                D_{p, \alpha}, &
 D_{{\bf n}, \alpha}^{\dagger} = {1\over\sqrt{N}}\suma_{p'}\e^{-\i{\bf p'
 \cdot n}} D_{p', \alpha}^{\dagger},
\ea
}
where {\bf p} is dimensionless momentum $(p_x,p_y) \in (0,2\pi)$
and taking into account the relation  ${1 \over N} \sum_{\bf n}\e^{\i({\bf
p}-{\bf p}')\cdot {\bf n}}  = \delta_{{\bf p}',{\bf p}}$ the Hamiltonian,
\Eref{TB}, is reduced to the form
$$
{\hat H}_{\rm TB}=\sum_{{\bf p}, \alpha} {\hat
 \psi}_{p,\alpha}^{\dagger} H_p {\hat \psi}_{p,\alpha},
$$
where
\eq{Hp}{
\psi_{p,\alpha} \equiv
\left[
 \ba{c}
  X_{p, \alpha} \\
  Y_{p, \alpha} \\
  S_{p, \alpha} \\
  D_{p, \alpha}
 \ea
 \right], \qquad
 H_p = \left( \ba{lccr}
   0        & -ts_{_{X}}s_{_{Y}} & I_s s_{_{X}}  & I_d s_{_{X}} \\
   -ts_{_{Y}}s_{_{X}} & 0        & I_s s_{_{Y}}  & -I_ds_{_{Y}} \\
   I_ss_{_{X}}   & I_s s_{_{Y}}  & \epsilon_s    & 0            \\
   I_d s_{_{X}}  & -I_d s_{_{Y}} & 0             &\epsilon_d
               \ea \right)
}
The notations used here are after Andersen \etal~\cite{Andersen}:
$s_{_X} = 2 \sin(p_x/2),$ $s_{_Y} = 2 \sin(p_y/2),$ $s =
(s^2_{_X}+s^2_{_Y})^{1 \over 2},$
$x \equiv (1-\cos p_x)/2$ and $y\equiv (1-\cos p_y)/2.$

To find the energy spectrum of the TB Hamiltonian one can employ the method
described in \Rref{Mishonov}. Its essence comprises in extracting an effective
oxygen part of the TB Hamiltonian by eliminating the metallic amplitudes
(a procedure also known as Loewdin perturbation technique). In our case
these are ${\tilde S}_p$ and ${\tilde D}_p$ which read as
\eq{SiD}{
{\tilde S}_p = -\frac{I_s}{\epsilon_s - \eps}(s_{_X} {\tilde X_p} +
s_{_{Y}} {\tilde Y_p}), \qquad
{\tilde D}_p =
-\frac{I_d}{\epsilon_d - \eps}(s_{_X} {\tilde X_p} - s_{_{Y}} {\tilde
Y_p}).
}
Hence we obtain $2 \times 2$ matrix problem.  The effective oxygen
$2 \times 2$ Hamiltonian  takes the form
$$
H^{\rm (O-O)}_{\rm eff}= H_0+V_t,
$$

\eq{Heff}{
H_0=-B_{\rm eff}
\left(
\ba{cc}
     s_{_{X}}s_{_{X}}   & s_{_{X}}s_{_{Y}} \\
  s_{_{Y}}s_{_{X}}      & s_{_{Y}}s_{_{Y}}
\ea
\right), \qquad
V_t=-t_{\rm eff}
\left(
\ba{lr}
   0          & s_{_{X}}s_{_{Y}} \\
  s_{_{Y}}s_{_{X}}      & 0
\ea
\right),
}
where $B_{\rm eff}=I_s^2/(\epsilon_s-\eps)+I_d^2/(\epsilon_d-\eps)$ and
$t_{\rm eff}=t-2I_d^2/(\epsilon_d-\eps).$

To solve the eigenvalue problem for $H^{\rm (O-O)}_{\rm eff}$ we
assume that $t_{\rm eff} \ll B_{\rm eff}$ and will use
perturbation theory with respect to the small parameter $\tau=t_{\rm
eff}/B_{\rm eff}.$ In zeroth order approximation we have
\eq{c0b0}{
\ba{ll}
\ec^{(0)}=0, & \vert c^{(0)} \rangle=\frac{1}{s}
\left(
\ba{c}
-s_{_{Y}}  \\
 s_{_{X}}
\ea
\right), \\ & \\
\eb^{(0)}=-B_{\rm eff}, & \vert b^{(0)} \rangle=\frac{1}{s}
\left(
\ba{c}
 s_{_{Y}}  \\
 s_{_{X}}
\ea
\right).
\ea
}
For our purposes we will consider the first order correction with respect
to $\vert c \rangle$ and $\ec$ They are given by (see for example
\Rref{LandauIII})
\eqn{
|c^{(1)}\rangle & = &\frac{\langle b^{(0)}|V_t|c^{(0)}\rangle \;
               |b^{(0)}\rangle}
               {\ec^{(0)}-\eb^{(0)}}, \label{c1} \\
\ec^{(1)}({\bf p}) & = &
               \langle c^{(0)}|V_t|c^{(0)}\rangle =
2 t_{\rm eff } \frac{s_{_{X}}^2 s_{_{Y}}^2}{s_{_{X}}^2+s_{_{Y}}^2}. \nn
}
One can readily obtain the required matrix element by using Eqs.~(\ref{Heff})
and~(\ref{c0b0})
$$
\langle b^{(0)}|V_t|c^{(0)}\rangle = -t_{\rm eff}
\frac{s_{_{X}}s_{_{Y}} \left(s_{_{X}}^2-s_{_{Y}}^2 \right)}
{s_{_{X}}^2+s_{_{Y}}^2}.
$$
Therefore, according \Eref{c1}, the first correction to the $\vert c\rangle$
vector takes the form
$$
|c^{(1)}\rangle =-\frac{\tau}{s^3}s_{_{X}}s_{_{Y}}
\left(s_{_{X}}^2-s_{_{Y}}^2\right)
\left(
\ba{c}
s_{_{X}} \\
s_{_{Y}}
\ea
\right).
$$
Now substituting \Eref{SiD}, for the conduction band in $(\tau \ll
1)$-approximation we finally get
\eq{cband}{
|c\rangle \simeq \frac{1}{s}
\left(
\ba{c}
-s_{_{Y}}+{-\tau \over s^2 }s_{_{X}}^2s_{_{Y}}(s_{_{X}}^2-s_{_{Y}}^2) \\ \\
 s_{_{X}}+{-\tau \over s^2 }s_{_{X}}s_{_{Y}}^2(s_{_{X}}^2-s_{_{Y}}^2) \\ \\
\frac{2I_s \tau }{\epsilon_s-\eps} s_{_{X}}s_{_{Y}}(s_{_{X}}^2-s_{_{Y}}^2) \\
                                                                           \\
\frac{I_d}{\epsilon_d-\eps} 2 s_{_{X}}s_{_{Y}}
  \left[ 1-\frac{\tau (s_{_{X}}^2-s_{_{Y}}^2)}{2s^2} \right]
\ea
\right),
}
\eq{ecband}{
\ec({\bf p})=4t_{\rm eff}\frac{1}{\frac{1}{2x}+\frac{1}{2y}}.
}
The last two expressions are used to derive in the next section the
four-fermion term which describes the interaction between electrons
leading to attraction.
%
\section{THE HEITLER-LONDON INTERACTION}                                     %
\label{sec:HL}                                                               %

In order to describe the effective interaction between electrons we shall start
here from two-electron exchange Hamiltonian. The underlying idea of a double
electron exchange has been considered, for example, in
Refs.~\cite{Zener,AndersonHasegawa,deGennes,Satpathy} and the original
Heitler-London's considerations in the theory of H$_2$ molecule consist in
involving a double electron exchange amplitude that takes into account the
correlated hopping between neighbouring atoms~\cite{HL}.

In the case of CuO$_2$ plane the transitions between Cu$4s,$ Cu$3d_{x^2-y^2}$
and O$2p_{\sigma}$ must be taken into account. The $2p_{\sigma}\leftrightarrow
4s$ transition amplitude is denoted by $J_{sp}$ in the following, and $J_{dp}$
stands for the $2p_{\sigma}\leftrightarrow 3d$ hopping, respectively.
In order to complete the investigation, started in
Refs.~\cite{Groshev,GroshevReview}, here we
will not take into account the O-O hopping amplitude. Consequently,  the four
fermion interaction reads as
\eqn{
H_{\rm HL}= -\frac{1}{2} \sum_{{\bf n},\alpha,\beta}
 & \left\{ J_{sp}^{} \right. & \left[
X^{\dagger}_{{\bf n},\beta}S^{\dagger}_{{\bf n},\alpha}X_{{\bf
n},\alpha}S_{{\bf n},\beta} +
Y^{\dagger}_{{\bf n},\beta}S^{\dagger}_{{\bf n},\alpha}Y_{{\bf
n},\alpha}S_{{\bf n},\beta} \right. \nn \\
& & \left. +
X^{\dagger}_{{\bf n},\beta}S^{\dagger}_{x+1,y,\alpha}X_{{\bf
n},\alpha}S_{x+1,y,\beta}
+ Y^{\dagger}_{{\bf n},\beta}S^{\dagger}_{x,y+1,\alpha}Y_{{\bf
n},\alpha}S_{x,y+1,\beta} \right] \label{HHL} \\
+ & J_{dp} & \left[
X^{\dagger}_{{\bf n},\beta}D^{\dagger}_{{\bf n},\alpha}X_{{\bf
n},\alpha}D_{{\bf n},\beta} +
Y^{\dagger}_{{\bf n},\beta}D^{\dagger}_{{\bf n},\alpha}Y_{{\bf
n},\alpha}D_{{\bf n},\beta}    \right. \nn \\
 & & \left.\left. +
X^{\dagger}_{{\bf n},\beta}D^{\dagger}_{x+1,y,\alpha}X_{{\bf
n},\alpha}D_{x+1,y,\beta}
+ Y^{\dagger}_{{\bf n},\beta}D^{\dagger}_{x,y+1,\alpha}Y_{{\bf
n},\alpha}D_{x,y+1,\beta} \right] \right\}, \nn
}
Each term could be compared to the corresponding one for H$_2$ molecule
in \Rref{GroshevReview}, $H_{\rm HL} \simeq \sum_{\alpha,\beta}J
a_{\alpha}^{\dagger}b_{\beta}^{\dagger}a_{\beta}b_{\alpha}.$

The direct substitution of the transformations below in the interaction
Hamiltonian
\eq{ampltds}{\ba{ll}
X_{{\bf n}, \alpha} = {1 \over \sqrt{2N}} \suma_{\bf p}\e^{\i{\bf p} \cdot {\bf n}}
\left(-\i\e^{\i p_x/2}\right)\left(\frac{-s_{_{Y}}}{s} \right) c_{p,\alpha}, &
X^{\dagger}_{{\bf n}, \beta} = {1 \over \sqrt{2N}}\suma_{\bf p'}\e^{-\i {\bf p'}
\cdot {\bf n}}\left(\i\e^{-\i p'_x/2}\right)\left(\frac{-s_{_{Y}}}{s} \right)
c^{\dagger}_{p',\beta}, \\ & \\
Y_{{\bf n}, \alpha} = {1 \over \sqrt{2N}}\suma_{\bf p}\e^{\i {\bf p} \cdot {\bf n}}
\left(-\i\e^{\i p_y/2}\right)\left(\frac{s_{_{X}}}{s} \right) c_{p,\alpha}, &
Y^{\dagger}_{{\bf n}, \beta} = {1 \over \sqrt{2N}}\suma_{\bf p'}\e^{-\i {\bf p'}
\cdot {\bf n}}\left(\i\e^{-\i p'_y/2}\right) \left(\frac{s_{_{X}}}{s} \right)
c^{\dagger}_{p',\beta}, \\ & \\
S_{{\bf n}, \beta} = { 1 \over \sqrt{2N}}\suma_{\bf q}\frac{2I_s \tau }{\epsilon_s}
\e^{\i{\bf q} \cdot {\bf n}}\frac{s_{_{X}}s_{_{Y}}(s_{_{X}}^2-s_{_{Y}}^2)}{s}
c_{q,\beta}, &
S^{\dagger}_{{\bf n}, \alpha} = {1 \over \sqrt{2N}}\suma_{\bf q'} \frac{2I_s \tau
}{\epsilon_s}\e^{-\i {\bf q'} \cdot {\bf n}}
\frac{s_{_{X}}s_{_{Y}}(s_{_{X}}^2-s_{_{Y}}^2)}{s} c^{\dagger}_{q',\alpha},\\ &
\ea
}
\eqn{
D_{{\bf n}, \beta} &=& {1 \over \sqrt{2N}}\suma_{\bf q} \frac{I_d
                 }{\epsilon_d-\eps}\e^{\i {\bf q} \cdot {\bf n}} 2
                 s_{_{X}}s_{_{Y}}\!\left[1-\frac{\tau
                 (s_{_{X}}^2-s_{_{Y}}^2)}{2s^2} \right]\! c_{q,\beta}, \nn \\
D^{\dagger}_{{\bf n}, \alpha} &=& {1 \over \sqrt{2N}}\suma_{\bf q'} \frac{I_d
                            }{\epsilon_d-\eps}\e^{-\i{\bf q'} \cdot {\bf n}}
                            2 s_{_{X}}s_{_{Y}}\!\left[1-\frac{\tau
                            (s_{_{X}}^2-s_{_{Y}}^2)}{2s^2} \right]\!
                            c^{\dagger}_{q',\alpha}, \nn
}
leads to the momentum representation of the interaction. Here we shall
suppose $\eps \ll \epsilon_d,$ and therefore ${\tilde
\epsilon}_d=\epsilon_d-\eps\simeq \epsilon_d.$ During the calculations we have
used the equality
$$ \frac{1}{N} \sum_{\bf n}\e^{-\i{\bf p' \cdot n}}\e^{-\i{\bf q' \cdot n}}
\e^{\i{\bf p\cdot n}}\e^{\i{\bf q\cdot n}} =\delta_{\bf p'+q', p+q},
$$
and thus we have a sum over four momenta which satisfies the quasimomentum
conservation law. In the case of space homogeneous order parameter and
currentless equilibrium state we must take into account only the terms with
zero momentum ${\bf p+q=p'+q'=}0,$ or ${\bf q=-p, q'=-p'},$ and this leads to
simplification of the result for $H_{\rm int}$
\eq{Hint}{
H_{\rm int}=-\frac{1}{2N} \sum_{{\bf p},{\bf p'},\alpha,\beta}
V({\bf p,p'}) c^{\dagger}_{p',\beta}c^{\dagger}_{-p',\alpha}
c_{p,\alpha}c_{-p,\beta},
}
where
\eqn{
V({\bf p',p}) & = &
\left\{ J_{sp}
        \left[
               \left( \frac{I_s}{\epsilon_s}
               \right)^2 \tau^2
                       \sigma({\bf p'}) \sigma({\bf p})
        \right]  \right.  \nn \\
& + & \left. J_{dp}
        \left[
               \left( \frac{I_d}{\epsilon_d}
               \right)^2
                      \frac{1}{s({\bf p'})}
               \left( 2-\tau \sigma^2({\bf p'})
               \right)
                      \frac{1}{s({\bf p})}
               \left( 2-\tau \sigma^2({\bf p})
               \right)
        \right]
\right\}  \label{Vcot} \\
 & \times &
\frac{s_{_{X}}^2({\bf p'})s_{_{Y}}^2({\bf p'})}{s({\bf p'})}
\frac{s_{_{X}}^2({\bf p})s_{_{Y}}^2({\bf p})}{s({\bf p})}
\left( 1+\frac{1}{2} \cot(p'_x/2)\cot(p_x/2)+
\frac{1}{2}\cot(p'_y/2)\cot(p_y/2) \right).   \nn
}
and
$$
\sigma({\bf p})=\frac{s_{_{X}}^2-s_{_{Y}}^2}{s}({\bf p}).
$$
We consider the case where the influence of the odd in $p_x$ and $p_y$ terms is
negligible. This, for example, holds for the
conventional superconductors, described by the BCS theory~\cite{BCS} and
could take place in the layered cuprates as shown by the experiments on
Bi$_2$SrCa$_2$Cu$_2$O$_8$~\cite{BSCCO}. In the next section we consider
the possibilities provided by the reduced kernel
\eqn{
V_{\rm HL}({\bf p',p}) & = &
\left\{
J_{sp}
\left(\frac{I_s}{\epsilon_s} \right)^2 \tau^2
\sigma({\bf p'}) \sigma({\bf p}) \right. \nn \\
 & + & \left.
J_{dp}\left[
\left(\frac{I_d}{\epsilon_d} \right)^2
\frac{1}{s({\bf p'})} \left( 2-\tau \sigma^2({\bf p'})\right)
\frac{1}{s({\bf p})} \left( 2-\tau \sigma^2({\bf p}) \right)
\right]
\right\} \label{Vnc} \\
 & \times &
\frac{s_{_{X}}^2({\bf p'})s_{_{Y}}^2({\bf p'})}{s({\bf p'})}
\frac{s_{_{X}}^2({\bf p})s_{_{Y}}^2({\bf p})}{s({\bf p})}. \nn
}
\section{THE BCS SCHEME}                                                     %
\label{sec:BCS}                                                              %

Consider now $V_{\rm HL}({\bf p',p})$ involved in the self-consistent BSC
calculation~\cite{BCS} of the superconducting gap.  Following the method
described in this fundamental work and notations from \Rref{LandauIX} we
obtain the following expression for the order parameter
\eq{BCSgap}{
\Delta({\bf p'})=\frac{1}{2N}\sum_{{\bf p} \in \cal L} V({\bf p',p})
\frac{\tanh\left(\frac{E({\bf p})}{2T}\right)}{E({\bf p})} \Delta({\bf p}),
\qquad
E({\bf p})=\sqrt{\Delta^2({\bf p})+\eta^2({\bf p})},
}
where $\eta({\bf p})={\bf p}^2/2m-\mu,$ with $\mu$ being the chemical potential
of the electrons. The renormalization procedure~\cite{LandauIX} tells us
that the summation in~\Eref{BCSgap} is only over a narrow energy interval along
the Fermi surface contour (FS $\cal L$), i.e. $E_{\rm F}-\hbar \omega_{_D}\leq
\eps({\bf p}) \leq E_{\rm F}+\hbar \omega_{_D},$
where the cut-off parameter of the sum $\hbar \omega_{_D}$ is found
to be $\hbar \omega_{_D}\approx E_{\rm F}/2.$ For more details about the
calculations see for example \Rref{Praga}, where the influence of the
impurities on the order parameter symmetry is studied as well. In the case
of layered cuprates the equation for the constant energy contours (CEC) is
given by
$\ec({\bf p})=E_{\rm F},$ where $\ec({\bf p})$ is given by \Eref{ecband}. The
result for $V_{\rm HL}$ can be further simplified if we introduce the
dimensionless Fermi energy measured in units of the conduction band width $w$
\eq{Fermi-line}{
\ew\equiv\frac{\ec({\bf p})}{w}=
\frac{1}{\frac{1}{2x}+\frac{1}{2y}}=
\frac{1}{2} \frac{s_{_{X}}^2s_{_{Y}}^2}{s_{_{X}}^2+s_{_{Y}}^2}
={\rm const},
}
where $w=4t_{\rm eff}$ is the bandwidth. Thus we get
\eqn{
V_{\rm HL}({\bf p',p}) & = & 16 \ew^2
\left\{
       J_{sp}
       \left( \frac{I_s}{\epsilon_s} \right)^2 \tau^2
       \left( s_{_{X}}^2({\bf p'})-s_{_{Y}}^2({\bf p'}) \right)
\left( s_{_{X}}^2({\bf p})-s_{_{Y}}^2({\bf p}) \right) \right. \label{Vinter}\\
 & + & \left. J_{dp}
\left[
       \left(\frac{I_d}{\epsilon_d} \right)^2
       \frac{1}{s({\bf p'})} \left( 2-\tau \sigma^2({\bf p'}) \right)
       \frac{1}{s({\bf p})}  \left( 2-\tau \sigma^2({\bf p})  \right)
\right] \right\}. \nn
}
\section{DISCUSSION}                                                        %
\label{sec:Discussion}                                                      %

To gain further knowledge on the gap symmetry it is straightforward to examine
\Eref{BCSgap} for particular choices of the interaction parameters entering
\Eref{Vinter} for which certain plausible limit cases occur.
Thus, for instance, if the $3d$ amplitudes dominate the transitions between
the Cu and O orbitals as considered in \Rref{Emery}, one would have $J_{sp}\ll
J_{dp}$ and therefore
\eqn{
V_{\rm HL}({\bf p',p}) & = & 16 \ew^2 J_{dp} \left[
       \left(\frac{I_d}{\epsilon_d} \right)^2
    \frac{1}{s({\bf p'})} \left( 2-\tau \sigma^2({\bf p'}) \right)
    \frac{1}{s({\bf p})} \left( 2-\tau \sigma^2({\bf p}) \right)
\right]  \nn \\
& = & 16 \ew^2 J_{dp}\left(\frac{I_d}{\epsilon_d} \right)^2 \chi_s({\bf
p'}) \chi_s({\bf p}), \nn
}
where
$$
\chi_s({\bf p})= \frac{1}{s(\bf p)}\left(
2 - \tau \frac{\left(s_{_X}^2({\bf p})-s_{_{Y}}^2({\bf p})\right)^2}
              {s_{_X}^2({\bf p})+s_{_Y}^2({\bf p})}\right).
$$
Thus we have a separable Hamiltonian of the interaction leading to an implicit
analytical solution for the gap
\eq{BCSsgap}{
\Delta({\bf p})=\frac{1}{2N}\sum_{{\bf p'} \in \cal L}
(16 J_{dp} \ew^2)\!\!
\left(\frac{I_d}{\epsilon_d}\right)^2 \chi_s({\bf p})\chi_s({\bf p'})
\frac{\tanh\left(\frac{E({\bf p'})}{2T}\right)}{E({\bf p'})} \Delta({\bf p'}).
}
After separating the angular dependence in $\chi_s({\bf p})$ and introducing
the so called {\it order parameter} $\Xi_s(T)$ at finite temperature $T,$ we
have
\eq{sum=1}{\ba{l}
\Delta({\bf p})  = \chi_s({\bf p})\,\Xi_s(T), \nn \\
\frac{1}{2N} 16 J_{dp}\ew^2
\left(
      \frac{I_d}{\epsilon_d}
\right)\!^2 \suma_{\bf p'} \chi_{s}^{2}({\bf p'})
\frac{\tanh\left(\frac{E({\bf p'})}{2T}\right)}{E({\bf p'})}  =  1,
\ea
}
where, in compliance with \Eref{BCSgap}
$$
E({\bf p})=\sqrt{\chi_{s}^{2}({\bf p})\Xi_{s}^{2}(T)+\eta^2({\bf p})}.
$$
This expression has the standard BCS form for a scalar type gap~\cite{BCS}.
In this case, $J_{sp}\ll J_{dp},$ we have the angular dependence
$$
  \Delta ({\bf p}) \propto \chi_s({\bf p}) > 0,
$$
and therefore, as we use $\tau \ll 1$, it exhibits $s$-type symmetry,
$\Delta \approx \mbox{const}.$ Such a possibility exists in strongly
irradiated samples~\cite{Radtke} (see also Refs.~\cite{Abrikosov,Pokrovsky}
and references therein) or $\NCCO.$

Consider now the opposite case $J_{dp}\ll J_{sp}.$ In the separable kernel
obtained
\eq{d-typeV}{
V_{\rm HL} \approx 16 \ew^2 J_{sp}
\left(\frac{I_s}{\epsilon_s} \right)^2  \tau^2
\chi_d({\bf p'})\chi_d({\bf p}),
}
the $\chi_d({\bf p})$ function has the form
$$
\chi_d({\bf p})=s_{_X}^2({\bf p})-s_{_Y}^2({\bf p})
= - 2 \left( \cos p_x - \cos p_y \right),
$$
which yields the so called $d$-type gap anisotropy. Now the gap equations
Eqs.~(\ref{BCSgap}), ~(\ref{BCSsgap}) and~(\ref{sum=1}) take the form
\eq{BCSdgap}{\ba{l}
\Delta({\bf p})=\chi_d({\bf p})\,\Xi_d(T) \propto \cos (p_x) - \cos (p_y),\\ \\
\frac{1}{2N} 16 J_{sp}\ew^2
\left(\frac{I_s}{\epsilon_s} \right)^2  \tau^2
\suma_{\bf p'} \chi_{d}^{2}({\bf p'})
\frac{\tanh\left( E({\bf p'})/2T \right)}{E({\bf p'})}=1,\\ \\
E({\bf p})=\sqrt{\chi_{d}^{2}({\bf p})\Xi_{d}^{2}(T)+\eta^2({\bf p})}.
\ea
}
Here the angular dependence is carried by the fragment $s_{_X}^2-s_{_Y}^2;$
the nodes of the gap are now situated at the points $p_x=\pm \pi \pm p_y,\;
p_x= \pm p_y,$ i.e. along the diagonals of the rounded FS square
$\ew=\ec({\bf p})/w=\mbox{const},$ in case of $\ew\leq 0.38$ as
it is for the hole doped YBa$_2$Cu$_3$O$_7$ and Bi$_2$Sr$_2$CaCu$_2$O$_8.$
Such a location of the nodes of $\Delta({\bf p})$ is in accordance with that
reported in \Rref{BSCCO} where at last the methodologically important
pair of parameters $(u,v)$ of the theory~\cite{LandauIX} have been measured by
ARPES.

\begin{figure}
\hspace*{0.2cm}\psfig{file=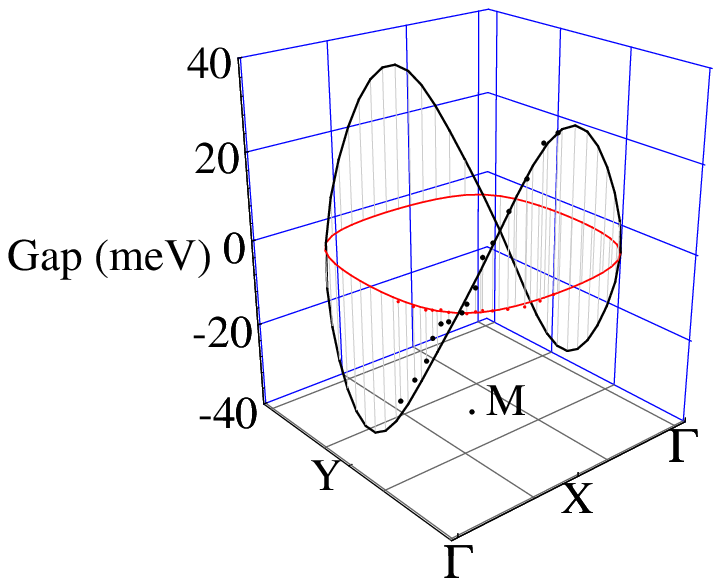,width=12cm}
\vskip1cm
\caption{The analytical '$\cos(p_x)-\cos(p_y)$' fit (thick solid line),
according to Eq.~(5.4),
to the experimental data by Dingh~\etal~[18] for the gap
anisotropy in Bi$_2$Sr$_2$CaCu$_2$O$_8$ (the dots). The fitted Fermi surface
line $\cal L$ (horizontal rounded square) is obtained by projecting 
the gap curve onto the $(p_x,p_y)$ plane. The notations for the 
high-symmetry points of the Brillouin zone are those standard for 
simple square lattice.}
\label{fig:fit}
\end{figure}

To bridge the current discussion and the experiment we employ \Eref{BCSdgap} to
fit the recent experimental data by Ding \etal~\cite{BSCCO} and the result is
shown in \Fref{fig:fit}. Since $\Delta({\bf p})$ lives only on the FS contour a
simultaneous fit to both the gap and FS can be achieved by simply projecting the
gap curve onto the $(p_x,p_y)$ plane. As clearly seen in \Fref{fig:fit}, the
adopted here simple TB self-consistent model remarkably reproduces the
experimentally observed $\Delta({\bf p})$ anisotropy in
Bi$_2$Sr$_2$CaCu$_2$O$_8.$

\section{CONCLUSIONS}                                                        %
\label{sec:Contusion}                                                        %

In the preceding sections we have given an account of a model in which
different cases of superconducting gap symmetry occur upon 'passing' various
sets of parameters as an 'input'.

Having obtained the underlying microscopic mechanism of high-$T_c$
superconductivity we hope that subsequent investigations on different materials
will finally determine the parameters entering the interaction Hamiltonian so
that its structure be enough to explain the experimentally observed different
types of superconductivity. Let us note that now in a decade of investigations
it is not yet firmly recognized whether it is a standard BCS scheme or some
kind of exotic interaction that gives rise to high-$T_c$ superconductivity.

In conclusion we stress that within the framework of the suggested model
not only the results that give $s$-type symmetry is easily interpreted,
but also the recent experiments on the Josephson $\pi$-shift in
YBa$_2$Cu$_3$O$_7$~\cite{Mathai} and the ARPES study of
Bi$_2$Sr$_2$CaCu$_2$O$_8$
and YBa$_2$Cu$_3$O$_7$~\cite{BSCCO,Aebi}. This model makes use of ideas having
their origin in the quantum chemistry, quantum field theory and gives, by
itself, a successive microscopic derivation of the interaction Hamiltonian
$H_{\rm int}$ of the BCS theory. Moreover the interpolation formulae used to
fit the experimental data for the Fermi contour and the angular dependence of
the order parameter, for instance, are obtained as a simple result within the
framework of the traditional band picture and the BCS scheme. We consider
that it is most unlikely the same analytic interpolation formulae to be
successively derived by an alternative theoretical model, i.e. model using only
college trigonometry and exhibiting textbook-like behaviour. In this sense the
theory of superconductivity repeats the development of quantum electrodynamics
from half century ago and we could see the victory of traditionalism in the
decadent theoretical physics at the end of the 20-th century.

\acknowledgments

This work was partially supported by the Bulgarian National Scientific Fund
under Contract No. Phys. 627/1996. The authors are much indebted to
Prof. I. Z. Kostadinov for allowing them unrestricted access to the computing
facilities of the Center for Space Research and Technologies, where the
essential part of this work was accomplished, and for the stimulating
discussions as well.

%

%
%
\end{document}